\documentclass[10pt,aps,prl,secnumarabic,superscriptaddress,nopacs,amsmath,amstex,amssymb,citeautoscript,longbibliography,floatfix,twocolumn]{revtex4-2}
\usepackage{graphicx,amsmath,amssymb,amsfonts,color}
\usepackage{nicefrac}
\usepackage{multirow, makecell, ragged2e}
\usepackage{physics}
\usepackage{bm}
\usepackage{float,lipsum}
\usepackage{algpseudocode}
\usepackage{algorithm}
\usepackage{placeins}

\usepackage[colorlinks,bookmarks=true,citecolor=blue,linkcolor=red,urlcolor=blue]{hyperref}

\usepackage[nameinlink,noabbrev]{cleveref}



\usepackage{graphicx}
\usepackage{subfigure}
\usepackage{adjustbox}
\usepackage{bm}
\usepackage{color}
\usepackage{braket}
\usepackage{standalone}
\usepackage{multirow}
\usepackage{tikz}
\usepackage{mathrsfs}
\usepackage{dsfont}
\usepackage{comment}
\usepackage{amsmath}

\usepackage{orcidlink}
\usepackage{subcaption}

\usepackage{graphicx,amsmath,amssymb,amsfonts, color}
\usepackage{nicefrac}
\usepackage{multirow, makecell, ragged2e}
\usepackage{bm}
\usepackage{float,lipsum}
\usepackage{algpseudocode}
\usepackage{algorithm}
\renewcommand{\vec}[1]{\mbox{\boldmath$#1$}}

\newcommand{\be}{\begin{equation}}
\newcommand{\ee}{\end{equation}}

\renewcommand{\vec}[1]{\bm{#1}}

\newcommand{\am}{\operatorname{am}}

\begin{document}

\title{Current Induced Switching of Superconducting Order and Enhancement of Superconducting Diode Efficiency}
\author{Uddalok Nag}
\affiliation{Department of Physics, 104 Davey Lab, The Pennsylvania State University, University Park, Pennsylvania 16802, USA}
\author{Jonathan Schirmer}
\affiliation{Department of Physics, William \& Mary, Williamsburg, Virginia 23187, USA}
\author{Chao-Xing Liu}
\affiliation{Department of Physics, 104 Davey Lab, The Pennsylvania State University, University Park, Pennsylvania 16802, USA}
\affiliation{Center for Theory of Emergent Quantum Matter, The Pennsylvania State University, University Park, Pennsylvania 16802, USA}
\author{J. K. Jain\orcidlink{000-0003-0082-5881}}
\affiliation{Department of Physics, 104 Davey Lab, The Pennsylvania State University, University Park, Pennsylvania 16802, USA}
\affiliation{Center for Theory of Emergent Quantum Matter, The Pennsylvania State University, University Park, Pennsylvania 16802, USA}

\begin{abstract}
We propose that the superconducting diode (SD) efficiency can be significantly enhanced near the transition between two superconducting states by choosing parameters where, before the system goes normal with increasing supercurrent, it switches into a different superconducting order for one direction of the current but not for the other. This mechanism for producing high SD efficiency relies on the expectation that the critical current depends sensitively on the superconducting order. We demonstrate this explicitly by performing detailed calculations for a bilayer superconductor with an in-plane magnetic field, which admits the standard Bardeen-Cooper-Schrieffer (BCS) and the orbital Fulde–Ferrell–Larkin–Ovchinnikov (FFLO) orders as a function of the strength of the magnetic field. We predict a sharp peak in the SD efficiency in the FFLO state close to the transition, which arises from a complex interplay between the two superconducting orders. An implication of our study is that the measurement of the SD efficiency can provide fundamental insight into the nature of the BCS-FFLO transition both as a function of the magnetic field and the supercurrent. 
\end{abstract}
\graphicspath{{./Figures/}}

\maketitle

{\it Introduction - } In the superconducting (SC) state, the condensation of Cooper pairs allows electric current to flow without energy dissipation, thereby producing a ``zero resistance" state that is a defining feature of superconductivity~\cite{Tinkham1996}. 
This supercurrent is the basis for a variety of potential applications of superconductors, including lossless power transmission~\cite{meyerhoff1973superconducting}, high-field SC magnets~\cite{avsner1999high}, quantum devices such as SQUIDs and SC qubits~\cite{yang2003}, and precision sensors for scientific and medical technologies~\cite{Danilin2024}. A superconductor 
undergoes a transition to the normal state when the supercurrent exceeds a critical value, a phenomenon often attributed to pair-breaking effects~\cite{Bagwell1994, Gurevich2007, Tinkham1996}. In conventional superconductors, the critical depairing current is typically reciprocal, that is, its magnitude remains the same for currents flowing in opposite directions. Recent theory and experiments~\cite{Cao2018,Ando2020,Davydova2022,Daido2022,He2022,Bauriedl2022, Yuan2022, Pal2022, Hou2023, Ghosh2024,  Ili2024,Hu2025, Chakraborty2025} 
have discovered nonreciprocity of dissipationless supercurrent in certain SC materials, where the supercurrent flows in one direction but not the opposite, a phenomenon called ``superconducting diode effect" in analogy to the nonreciprocal transport in a semiconducting diode. The superconducting diode (SD) efficiency is defined as
\begin{equation}
    \label{eta}
    \eta=|I_c^+-I_c^-|/(I_c^++I_c^-) \, \text{,}
\end{equation}
where $I_c^{\pm}$ are the critical currents in two opposite directions. Similarly to the central role of semiconducting diodes in modern electronic technologies, the SD effect may have potential device applications in superconducting rectifiers \cite{Ingla-Ayns2025} and logic circuits \cite{hosur2024}, cryogenic computing, and low-loss energy conversion systems \cite{Ma2025, Ando2020}.

The objective of the present work is to propose a mechanism to enhance the SD efficiency near the transition between two distinct SC states, denoted as SC$_1$ and SC$_2$, in the parameter regime when a SC$_1$-normal transition occurs as the current increases in one direction, but a SC$_1$-SC$_2$-normal transition occurs in the opposite direction.
The enhancement of SD efficiency occurs because of the critical current depends sensitively on the superconducting order. 
As an explicit demonstration, we consider a bilayer superconductor with an in-plane magnetic field, which has been shown to admit two types of SC orders.
At low magnetic fields the phases of the order parameters in two layers are equal, which we refer to as the Bardeen-Cooper-Schrieffer (BCS) state. As the magnetic field is increased, its coupling to the electron motion induces a transition into an {\em orbital} Fulde–Ferrell–Larkin–Ovchinnikov (FFLO) state with finite momentum pairing, as was predicted by Ref.~\cite{Liu17}, where a spatially varying phase difference exists between two layers~\cite{Liu17}. Evidence for this state has been revealed by probing rotational symmetry breaking in the SC state~\cite{Wan2023, Zhao2023}. Further theoretical work has revealed the important role of interlayer vortices in the transition regime from the BCS state to the orbital FFLO state~\cite{Nag2025}. (We neglect throughout this paper the Zeeman coupling of magnetic field to the electron spin, as is valid for the class of non-centrosymmetric layered superconductors known as Ising superconductors; here, strong Ising SOC pins electron spin into the out-of-plane direction and thus suppresses the Zeeman effect of in-plane magnetic fields, leading to a high upper critical field several times above the Pauli limit~\cite{Lu2015,Saito2016,Xi2016,Wan2023}.)

It is well known that increasing the supercurrent through a superconductor drives a transition into the normal state beyond a critical current. We show below that, for magnetic fields when two types of superconducting orders are in close competition, it is possible that increasing the supercurrent in one direction switches FFLO into BCS before the system eventually becomes normal, but in the opposite direction, the transition happens from FFLO to normal directly. We find a marked enhancement of the SD efficiency in this regime.  

We emphasize that the enhancement mechanism of SD effect considered here differs from that associated with the FFLO state discussed in Ref.~\cite{Yuan2022,Pal2022, Hou2023, Chakraborty2025,Bhowmik2025}.
These consider a class of non-centrosymmetric superconductors, where the interplay between the Rashba spin-orbit coupling (which produces spin-textured Fermi surfaces) and the Zeeman splitting from in-plane magnetic fields can drive
a spatial modulation of the {\it overall} SC phase, yielding the so-called helical SC state or single-$\bf{q}$ FFLO state~\cite{Kaur2005, Yuan2022, Smidman2017, Samokhin2014}. The momentum ${\bf q}$ in the FFLO imposes a directional asymmetry in the critical current, which naturally results in SD effect~\cite{Yuan2022}.
In contrast, the {\it orbital} FFLO state in a SC bilayer considered here involves a {\it relative} phase shift of SC order parameters between the two layers rather than an overall phase modulation, and hence does not by itself produce an SD effect. In our case, the enhancement of SD effect near the BCS–FFLO transition arises from the coupling between interlayer vortices and the net supercurrent when inversion symmetry between two layers is broken.

{\it  Phase diagram as a function of current and magnetic field -- } We now describe how we obtain the phase diagram of a SC bilayer in the presence of a supercurrent and an in-plane magnetic field by solving the Ginzburg–Landau (GL) equations self-consistently. We denote the SC order parameters in the layer $l=1,2$ as $\Psi_l = \psi_l e^{i(\xi(x)-(-1)^l\phi(x)/2+qx)}$, 
where the amplitude $\psi_l$ is taken to be spatially uniform (previous calculations have shown a very weak dependence on $x$ in the FFLO state~\cite{Nag2025}), $\xi(x)$ and $\phi(x)$ are periodic over the length $L_x$ of the system and describe the overall and relative phases between two layers. We also add the phase $e^{iqx}$ to control the current through the system. Based on the order parameter form, the Ginzburg-Landau (GL) free energy is written as

\begin{widetext}
\begin{eqnarray} \label{fe1} 
    \Omega[\Psi_l] 
     & = &\frac{1}{V_0}\int d^2\vec{r} \left[\sum_l \alpha_l\psi_l^2 + \frac{\beta}{2}\sum_l\psi_l^4
    +\frac{\psi_1^2}{2m}\left(q-q_{_{B}}+\frac{\nabla\phi}{2}\right)^2 + \frac{\psi_2^2}{2m}\left(q+q_{_{B}} - \frac{\nabla\phi}{2}\right)^2 
     -2J\psi_1\psi_2 \cos \phi \right] , \label{fe2}
\end{eqnarray}
\end{widetext}
where $\alpha_l$ controls the onset of superconductivity in layer $l$ and changes sign at the mean field transition temperature ($\alpha_l \propto (T-T_{c,l})$), the fourth order term $\beta>0$ to ensure that the system is stable, $m$ is the cooper pair mass and $J$ is the strength of the interlayer Josephson coupling. We allow $\alpha_1 \neq \alpha_2$ to break the inversion symmetry between two layers. To describe an in-plane magnetic field $\vec{B} = B\hat{e}_y$, we also include the vector potential $\vec{A}_l = (-1)^l \frac{d}{2}\vec{B} \times\hat{e}_z = (-1)^l \frac{dB}{2}\hat{e}_x$, where $d$ is the separation between the two layers, and introduce the parameter $q_{_{B}} = edB=2\pi\Phi/(L_x \Phi_0)$, where $\Phi$ is the net flux between two layers in the system of length $L_x$ and $\Phi_0 = {hc}/{e}$ is the flux quantum. For a given $q_{_{B}}$, the density of magnetic flux quanta is given by $\Phi/(L_x\Phi_0)=q_{_{B}}/(2\pi)$. We consider periodic boundary conditions in the $x$ direction. The periodic overall phase $\xi(x)$ can be gauged away by making the replacement $A_l\rightarrow A^\prime_l =A_l-\frac{1}{2e}\nabla\xi(x)$ for both layers.

We need to minimize the GL free energy with respect to $\psi_1$, $\psi_2$ and $\phi(x)$ for fixed $q$ and $q_{_{B}}$, and we refer to the minimum energy as the condensation energy $E_{\rm{con}}(q,q_{_{B}})$.
Variation with respect to $\phi(x)$ produces the sine-Gordon (SG) equation  
\begin{equation}
\label{SG}
    \partial_w^2\phi = \sin\phi 
\end{equation}
where $w = (4J^\prime m^\prime)^{1/2}x$ is a rescaled length, with $m^\prime =2m/(\psi_1^2+\psi_2^2)$ and $J^\prime =\psi_1\psi_2J$. As discussed in Sec. (S2) of Supplementary Material (SM)~\cite{SM_Nag}, the solutions of the SG equation are given by
\begin{equation}
    \label{am}
    \phi = 2\am\Big(\pm\frac{1}{k}w,k^2 \Big) - \pi,
\end{equation}
where $\am$ is the Jacobi amplitude function~\cite{abramowitz1965handbook, Scott1973}. 
The parameter $0< k\leq 1$ controls the density of vortices per unit length, $\rho = \sqrt{J^\prime m^\prime}/(kK(k^2))$, where $K$ is the complete elliptic integral of the first kind~\cite{abramowitz1965handbook, Scott1973} and a single vortex refers to a rotation of $\phi$ by $2\pi$.
The trivial phase-locked solution $\phi = 0$ ($\rho=0$) is obtained for $k=1$, as $K(k^2)$ diverges logarithmically as $k\rightarrow 1$; this will be referred to as the BCS state. The state in which $\phi$ varies with $x$ contains interlayer vortices and corresponds to the (orbital) FFLO state~\cite{Liu17, Wan2023}.

\begin{figure*}
    \centering
    \captionsetup{justification=raggedright,singlelinecheck=false}
    \includegraphics[width=1.0\linewidth]{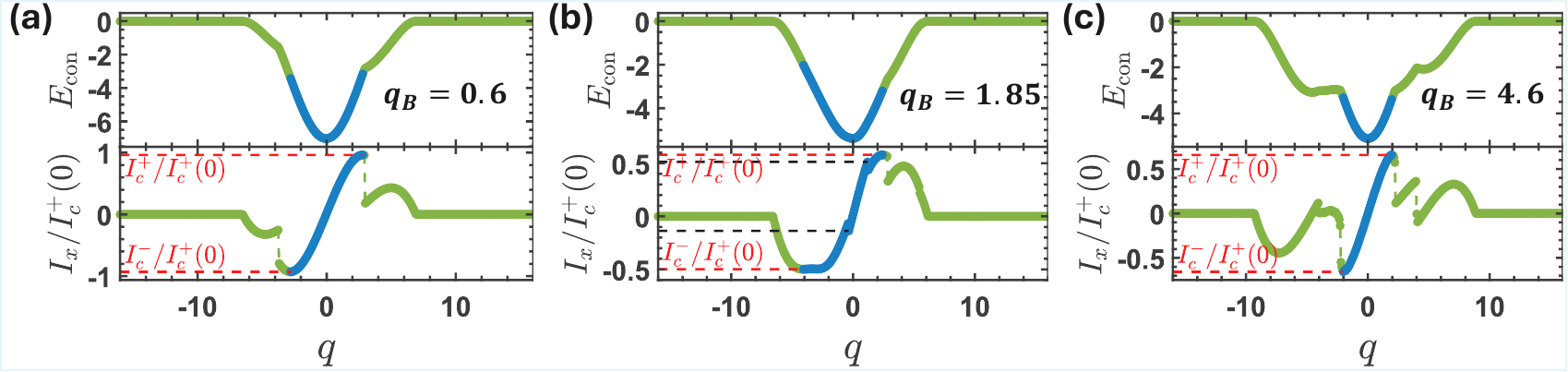}
    \caption{{\it Upper panels: } Ground state energy $E_{\rm con}$ versus momentum $q$. The results are shown for a bilayer with asymmetry parameter $a=0.1$ at three magnetic field ($q_B$) values.  {\it Lower panels:} Supercurrent $I_x= 2e\partial E_{\rm con}/\partial q$ versus $q$. When a given $I_x$ is produced by more than one value of $q$, we choose $q$ corresponding to the lowest energy, highlighted in blue. Critical currents $I_c^+$ and $I_c^-$ (marked in red) correspond to the maximum supercurrent. Black dashed lines mark phase boundaries between FFLO and BCS orders. (a) $q_{_{B}} = 0.6$: The system remains in the BCS state for all current values. (b) $q_{_{B}} = 1.85$: At zero current, the system is in the FFLO state. Increasing current in either direction induces a transition into the BCS state (crossing the black dashed lines) before ultimately driving the system normal. (c) $q_{_{B}} = 4.2$: The system remains in the FFLO state for all current values.}
    \label{fig:groundstate}
\end{figure*}

\begin{figure*}[t]
\captionsetup{justification=raggedright,singlelinecheck=false}
    \includegraphics[width=0.9\textwidth]{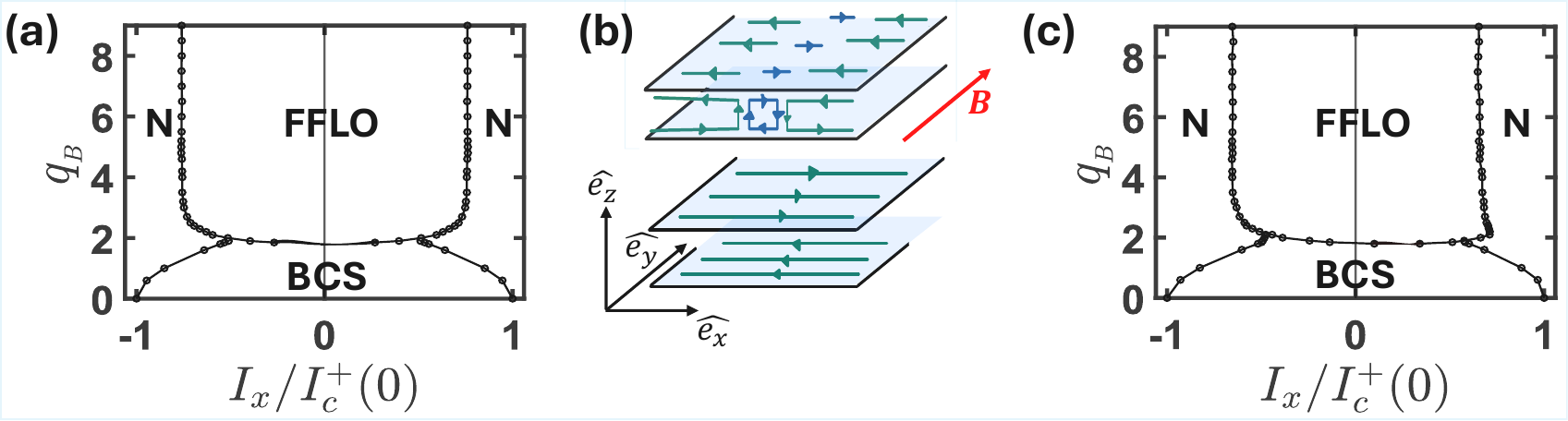}
    \caption{
    (a) Phase diagram of the symmetric bilayer ($\alpha_1 = \alpha_2 = -10$) as a function of the current $I_x$ (in units of $I_c^+(0)$, the critical current at zero magnetic field) and $q_B\propto B$ ($q_B$ is $2\pi$ times the number of magnetic flux quanta per unit length). The BCS and FFLO states separated by second-order transitions, and by first-order transitions from the normal (N) state at high currents. (b) Schematic illustration of the BCS state with uniform interlayer phase ($\phi = 0, \rho=0$, bottom) and FFLO state with spatially varying interlayer phase ($\phi \neq 0$, $\rho\neq 0$,  top). Green/blue arrows indicate local current patterns at $I_x = 0$. The in-plane magnetic field is $\vec{B} = B\hat{e}_y$. (c) Phase diagram of an asymmetric bilayer ($\alpha_1 = -9$, $\alpha_2 = -11$), which does not have inversion symmetry. The phase boundaries are asymmetric about $I_x = 0$, yielding unequal critical currents $I_c^+$ and $I_c^-$ and thus a nonzero superconducting diode efficiency. For both phase diagrams: $\beta = 20$, $J = 2$, $m = 1$ (these parameters are defined in main text and related to microscopic quantities in in Sec.~(S1) of the SM) }
    \label{fig:pd}
\end{figure*}

We next substitute Eq.~(\ref{am}) into the expression for the GL free energy $\Omega$ in Eq.(\ref{fe2}) and minimize it with respect to $\psi_1$, $\psi_2$ and $k$ ($k$ fixes $\phi$) numerically for given $q$ and $q_{_{B}}$ to get $E_{\rm{con}}(q,q_{{_B}})$, shown in the upper panels of Fig.~\ref{fig:groundstate} for different values of $q_{B}$. The supercurrent is given by $I_x=2e(\partial\Omega/\partial q)_{\rm min}=2e\partial E_{\rm con}/\partial q$,
\begin{align}
    \label{current_in_layer}
    I_x = \frac{2e}{m}\big[(\psi_1^2+\psi_2^2)q-(\psi_1^2-\psi_2^2)(q_{_{B}}-\nabla\phi/2))\big]_{\rm min} \,,
\end{align}
where the subscript min implies evaluation at values of $\psi_l$ and $\phi$ that minimize the free energy.  [In all the figures shown, we normalize the value of the current by the critical current at zero magnetic field, $I_c^+(0)$; at $q_{_{B}}=0$, the critical current is same in both directions due to time-reversal symmetry]. The supercurrent is shown in the lower panels of Fig.~\ref{fig:groundstate}. In cases when multiple values of $q$ yield the same current, we select the $q$ (highlighted in blue) with the lowest energy. 
The critical currents $I_c^+$ and $I_c^-$ can be read off directly from Fig.~\ref{fig:groundstate}. 

We display the phase diagram in the $q_{_{B}}$-$I_x$ plane in Fig.~\ref{fig:pd}. Panel (a) shows the phase diagram for a symmetric bilayer ($\alpha_1=\alpha_2$), while panel (c) for an asymmetric bilayer with the asymmetry factor 
$a=|(\alpha_1-\alpha_2)/(\alpha_1+\alpha_2)|=0.1$. At small $B$, the system is in the BCS state with no interlayer vortices ($\phi=0$), but it makes a transition with increasing $B$ into the FFLO state with interlayer vortices.

\begin{figure}[H]
\centering
\captionsetup{justification=raggedright,singlelinecheck=false}
    \includegraphics[width=0.48\textwidth]{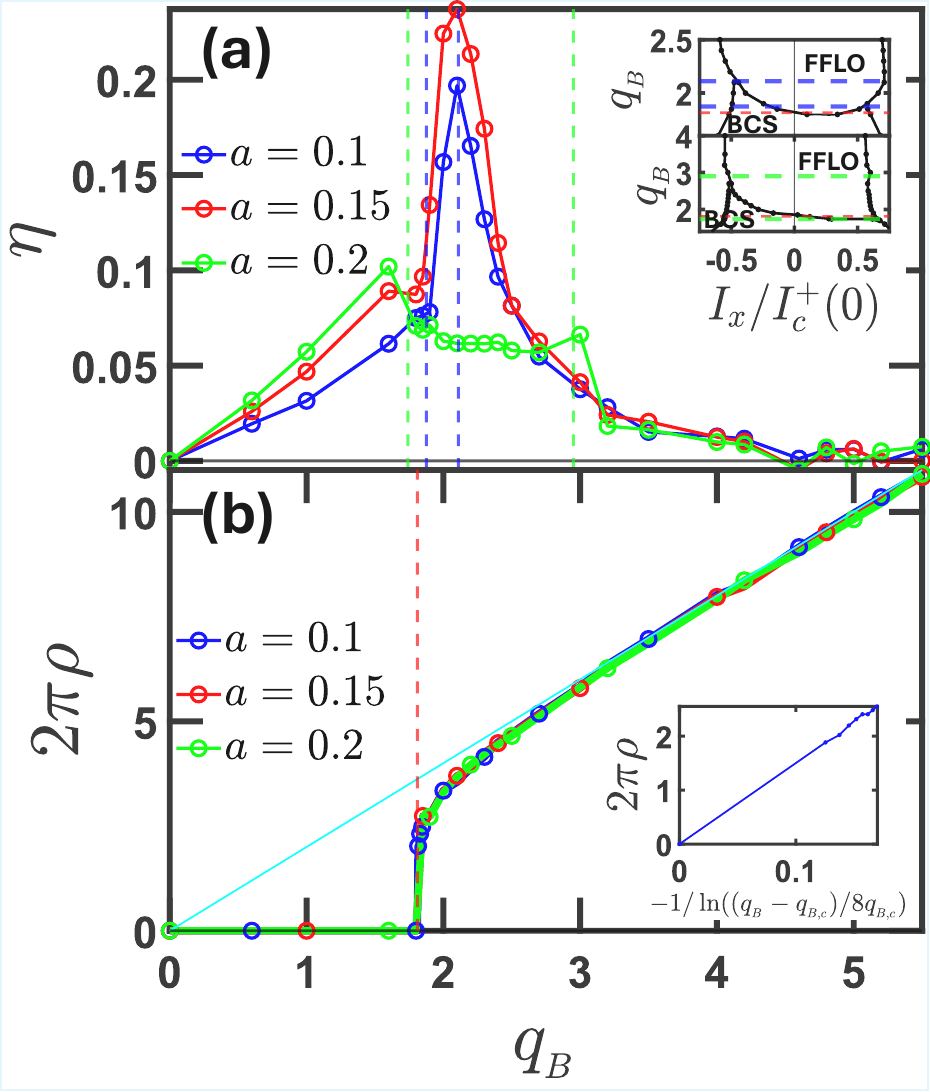}\hfill
    \caption{ 
    (a) Superconducting diode efficiency $\eta$ as a function of magnetic field $q_{_{B}}$ for various values of asymmetry parameter $a=|(\alpha_1-\alpha_2)/(\alpha_1+\alpha_2)|$. The upper (lower) inset shows magnified phase diagram from Fig.~\ref{fig:pd}(c) at $a=0.1$ ($a=0.2$). In both insets, the region between the horizontal dashed lines (blue for $a=0.1$, green for $a=0.2$) marks the $q_{_{B}}$ range where current in the $+x$ direction induces a FFLO$\rightarrow$N transition while current in the $-x$ direction induces a FFLO$\rightarrow$BCS$\rightarrow$N transition. These lines correspond to the color-matched vertical dashed lines in the main panel, marking the region of enhanced $\eta$. (b) Interlayer vortex density $\rho$ as a function of magnetic field $q_{_{B}}$ at $I_x=0$. The light blue line shows the high-field limiting case $\rho = 2q_{_{B}}$, corresponding to maximum vortex density where the layers effectively decouple. The vertical red dashed line marks the value of $q_{_{B}}$ at which vortices start entering the system at zero current. The inset depicts the vortex density $\rho$ plotted as a function of $-1/\ln[(q_B-q_{B,c})/8q_{B,c}]$, in accordance with the asymptotic behavior derived in Eq.~(S84) in Sec.~(S6) of the SM for $a=0.1$, to make it evident that the vortices enter the system continuously.}
    \label{fig:eta}
\end{figure}

At fixed $B$ (fixed $q_{_{B}}$), as the current $I_x$ exceeds a critical value, there is a transition into the normal state.
Interestingly, however, for certain values of $q_{_{B}}$, the system can switch from FFLO to BCS as the current is increased, before being driven normal. Even re-entrant transitions like BCS-FFLO-BCS-normal are possible [see the inset of Fig.~\ref{fig:eta}(a) which is zoomed-in region of Fig.~\ref{fig:pd}(c)]. Such current induced switching of the superconducting order lies at the root of the enhancement of SD efficiency, as discussed below. 

{\it Superconducting diode efficiency - } The SD efficiency $\eta$, defined in Eq.~(\ref{eta}), is zero whenever the second term inside the square brackets in Eq.(\ref{current_in_layer}), which describes the Meissner current, is zero. That, in turn, is the case when (i) the system has inversion symmetry ($\psi_1=\psi_2$); or (ii) $q_{_{B}}=0$; or (iii) the Abrikosov vortex density is $\rho = 2q_{{_B}}/(2\pi)$, i.e. equal to twice the number of magnetic flux quanta passing through the layers.
In general, the evaluation of $\eta$ requires a detailed calculation. Fig.~\ref{fig:eta} shows the SD efficiency $\eta$ as a function of the magnetic field for several values of the asymmetry parameter $a$.  

For small $q_{_{B}}$, the system is in the BCS state with $\phi=0$ and we note a linear increase of $\eta$ as a function of $q_{_{B}}$. This follows from Eq.~(\ref{current_in_layer}) with $\phi=0$, provided that $\psi_1$ and $\psi_2$ are independent of $q_{_{B}}$ to leading order, which is shown to be the case in Sec.~(S3) of the SM. 

As a reference, let us take the model considered in Ref.~\cite{Liu17} which studied the BCS-FFLO transition by assuming that the vortex density in the FFLO state is given by twice the magnetic flux density. In this model, $\eta$ increases with $q_{_{B}}$ in the BCS state, but  drops discontinuously to zero as soon as the system makes a transition into the FFLO state.

Our results for $\eta$ are qualitatively different. Remarkably, for small $a$, $\eta$ actually increases rapidly, rather than dropping to zero, as the system enters the FFLO state, but at some point it begins to decrease rapidly, approaching zero at large $q_{_{B}}$. The most striking feature is a sharp peak in $\eta$ on the FFLO side of the transition. In a broad sense, this physics arises only with the more detailed treatment of the BCS-FFLO transition in terms of the sine-Gordon model, wherein the vortex density rises continuously from zero to twice the magnetic flux density [see Fig.~\ref{fig:eta} (b) and its inset and Sec.~(S6) of SM]. Furthermore, the enhancement can be seen to occur in the region where the superconducting order switches from FFLO to BCS as the supercurrent is increased in one direction while remains in the FFLO in the other. In the insets of Fig.~\ref{fig:eta}(a), this region is marked between two horizontal dashed lines, blue (green) for $a=0.1$ ($a=0.2$), which correspond to the vertical dashed lines in the main figure. We attribute the enhanced SD efficiency to the sensitive dependence of the supercurrent to the type of superconducting order, which enhances the asymmetry in this region. 

It is intuitively expected that there are two vortices per magnetic flux quantum in the limit $q_B\rightarrow \infty$, implying that $\eta$ vanishes here. Mathematically, the limit $q_{{_B}}\rightarrow \infty$ can be shown to be equivalent to $J\rightarrow 0$ (see Sec.~(S5) of SM), where the two layers effectively decouple. For $J=0$, minimization with respect to $\phi$ yields $\nabla \phi(x) \rightarrow 2 q_{_{B}}x$, which corresponds to the vortex density $\rho = 2q_{{_B}}/(2\pi)$, which is equal to twice the number of magnetic flux quanta passing through the layers. With this form of $\phi$, and with $J=0$, $\Omega$ is an even function of $q$, i.e. the current is an odd function of $q$, implying $\eta=0$. For details, see Sec.~(S4) of the SM where we prove up to the order of $1/q_{_{B}}^3$.

{\it Conclusion and Discussion - } 
In summary, we have demonstrated that a supercurrent can induce a BCS-FFLO transition in a SC bilayer under in-plane magnetic fields. This current-induced transition is asymmetric with respect to two current directions, leading to an enhancement of SD efficiency. 

Our mechanism relies on interlayer vortices generated by the orbital effect of in-plane magnetic field in combination with interlayer Josephson coupling, and thus does not, in principle, require spin-orbit coupling.  It is distinct from the existing theories of SD effect relying on strong spin-orbit coupling and Zeeman effects~\cite{Yuan2022,Chakraborty2025} and thus broadens the range of systems in which an enhanced SD efficiency may be realized. Our prediction are directly relevant to recent experiments on layered superconductors with orbital FFLO state~\cite{Sohn2018, Barrera2018, Bauriedl2022, Wan2023, Zhao2023}.

An important implication of our work is that the measurement of the SD efficiency provides fundamental insight, and can thus serve as a probe into the nature of the phase transition. As noted above, a sharp peak in $\eta$ on the FFLO side is an indication of the continuous nature of the BCS-FFLO transition, as well as of a supercurrent induced FFLO-BCS transition. Finally, we expect similar enhancement of the SD efficiency in other systems with competing superconducting orders, but leave detailed studies for future work.

{\it Acknowledgments -} We gratefully acknowledge Dr. Debmalya Chakraborty for insightful discussions. UN and JKJ were supported in part by the U. S. Department of Energy, Office of Basic Energy Sciences, under Grant no. DE-SC0005042. JS was supported by the U.S. US Department of Energy, Office of Basic Energy Sciences, via Award No. DE-SC0022245. C.-X.L 
acknowledges the support from the NSF through The Pennsylvania State University Materials Research Science and Engineering Center [DMR-2011839]. We acknowledge Advanced CyberInfrastructure computational resources provided by The Institute for CyberScience at The Pennsylvania State University.

\bibliographystyle{unsrtnat}
\bibliography{Schirmer.bib}

\end{document}


\title{Supplemental Material for ``Current Induced Switching of Superconducting Order and Enhancement of Superconducting Diode Efficiency"}
\author{Uddalok Nag}
\affiliation{Department of Physics, 104 Davey Lab, The Pennsylvania State University, University Park, Pennsylvania 16802, USA}
\author{Jonathan Schirmer}
\affiliation{Department of Physics, William \& Mary, Williamsburg, Virginia 23187, USA}
\author{Chao-Xing Liu}
\affiliation{Department of Physics, 104 Davey Lab, The Pennsylvania State University, University Park, Pennsylvania 16802, USA}
\affiliation{Center for Theory of Emergent Quantum Matter, The Pennsylvania State University, University Park, Pennsylvania 16802, USA}
\author{J. K. Jain\orcidlink{000-0003-0082-5881}}
\affiliation{Department of Physics, 104 Davey Lab, The Pennsylvania State University, University Park, Pennsylvania 16802, USA}
\affiliation{Center for Theory of Emergent Quantum Matter, The Pennsylvania State University, University Park, Pennsylvania 16802, USA}

\maketitle

\setcounter{equation}{0}
\setcounter{figure}{0}
\setcounter{table}{0}
\setcounter{page}{1}
\setcounter{section}{0}

\makeatletter
\renewcommand{\theequation}{S\arabic{equation}}
\renewcommand{\thefigure}{S\arabic{figure}}
\renewcommand{\thetable}{S\arabic{table}}
\renewcommand{\thepage}{\arabic{page}}

\setcounter{secnumdepth}{2} 

\renewcommand{\thesection}{S\arabic{section}}
\renewcommand{\thesubsection}{\thesection.\arabic{subsection}}

\renewcommand{\@seccntformat}[1]{\csname the#1\endcsname:\quad}
\makeatother

\crefname{section}{}{} 
\Crefname{section}{}{} 
\crefformat{section}{#2#1#3} 

In this Supplementary Material, we provide the technical details and derivations supporting the results presented in the main text. 
In Sec.~(\cref{app:lattice2GL}), we derive the Ginzburg-Landau (GL) free energy from a microscopic bilayer negative-$U$ Hubbard model using the Hubbard-Stratonovich transformation.
The minimization of this free energy is discussed in Sec.~(\cref{app:SG}), where we map the problem onto the sine-Gordon equation and obtain solutions in terms of Jacobi elliptic functions. 
The subsequent sections explore specific limits of the theory: Sec.~(\cref{app:lowmag}) analyzes the low-magnetic-field regime to establish the linear dependence of the diode efficiency on the magnetic field ($\propto q_{_{B}}$), while Sec.~(\cref{app:highmag1}) and Sec.~(\cref{app:highmag}) examine the high-field limit to demonstrate how reciprocity is recovered as the layers effectively decouple. 
Finally, Sec.~(\cref{app:asymptotic}) provides an analytical derivation of the interlayer vortex density in both the sparse and dense vortex limits, the former confirming the continuous nature of the BCS-FFLO transition.

\vspace{2em} 

\section{The Ginzburg-Landau theory of bilayer superconductor}

\label{app:lattice2GL}
We start with a bilayer lattice model where each layer is described by a negative U Hubbard model
\begin{align}
    \mathcal{H}_l = &-t\sum_{<i,j>,\sigma}\hat{c}^\dagger_{l,i,\sigma}\hat{c}_{l,j,\sigma}-\mu\sum_{i,\sigma}\hat{c}^\dagger_{l,i,\sigma}\hat{c}_{l,i,\sigma}\\ \nonumber &-U\sum_i\hat{c}^\dagger_{l,i,\uparrow}\hat{c}^\dagger_{l,i,\downarrow}\hat{c}_{l,i,\downarrow}\hat{c}_{l,i,\uparrow} \, \text{,}
\end{align}
with $U>0$. The nearest neighbour interlayer coupling is given by
\begin{align}
    \mathcal{H}_\perp = -t_\perp \sum_{i,\sigma}\big( \hat{c}^\dagger_{1,i,\sigma}\hat{c}_{2,i,\sigma} + h.c.\big). 
\end{align}
and the full Hamiltonian is given by
$$\mathcal{H} = \mathcal{H}_1 + \mathcal{H}_2 + \mathcal{H}_\perp\, \text{.}$$ 
The partition function is written as
\begin{equation}
    \mathcal{Z} = \mathrm{Tr}\big(e^{-\beta_T\mathcal{H}} \big) \, \text{,}
\end{equation}
where $\beta_T = 1/T$ and $T$ is the temperature. We introduce Grassmannian fields $c$ and $c^*$ defined on each $\{l,i,\sigma\}$. The partition function can then be written as a path integral form
\begin{align}
    \mathcal{Z} = \int \mathcal{D}[c^*,c] e^{-S[c^*,c]}
\end{align}
with the action is given by
\begin{align}
    S = \int_0^{\beta_T} d\tau \Big[ \sum_{l,i,\sigma} c^*_{l,i,\sigma}(\partial_\tau-\mu)c_{l,i,\sigma} + \mathcal{H}[c^*,c] \Big].
    \label{action}
\end{align}
Note that the Grassmannian fields are antiperiodic in imaginary time with the period $\beta_T$,
$$c_{l,i,\sigma}(\beta_T) = -c_{l,i,\sigma}(0) \; \; \; c^*_{l,i,\sigma}(\beta_T) = -c^*_{l,i,\sigma}(0). $$ 

For the quartic terms, we perform the Hubbard-Stratonovich transformation by introducing field $\Psi_{l,i}(\tau)$, which leads to
\begin{align}
    e^{U\int d\tau c^*_\uparrow c^*_\downarrow c_\downarrow c_\uparrow}& = \int \mathcal{D}[\Psi,\Psi^*] \nonumber \\ &\exp\Big[-\int d\tau \Big( \frac{|\Psi|^2}{U} + \Psi c^*_\uparrow c^*_\downarrow + \Psi^* c_\downarrow c_\uparrow  \Big) \Big]. 
\end{align}

Thus, our partition function is expressed in terms of a path integral over both the Grassmannian fields and the new fields ($\Psi$) we introduced. Our strategy is to carry out the integration over the former to get an effective action in terms of the latter. We assume $\Psi$ is uniform in space and imaginary time: $\Psi_{l,i}(\tau) \rightarrow \Psi_l$. We will latter allow $\Psi$ to vary slowly to compute the kinetic energy term of the GL free energy.

The action for each layer becomes
\begin{align}
    S^\prime_l = \int_0^{\beta_T} & d\tau \Big[ \sum_{l,i,\sigma} c^*_{l,i,\sigma}(\partial_\tau-\mu)c_{l,i,\sigma} - t \sum_{<i,j>,\sigma}c^*_{l,i,\sigma}c_{l,j,\sigma}\nonumber \\ & +\sum_i\Big( \frac{|\Psi_l|^2}{U} + \Psi_l c^*_{l,i,\uparrow} c^*_{l,i,\downarrow} + \Psi_l^* c_{l,i,\downarrow} c_{l,i,\uparrow} \Big) \Big]. 
\end{align}
We also have the part of the action for interlayer tunneling
\begin{align}
    S^\prime_\perp &= \int_0^{\beta_T} d\tau \mathcal{H}_\perp [c^*, c] \nonumber \\
            &=\int_0^{\beta_T} d\tau\Big[ -t_\perp\sum_{i,\sigma}\Big( c^*_{1,i,\sigma} c_{2,i,\sigma} + c^*_{2,i,\sigma} c_{1,i,\sigma} \Big) \Big]. 
\end{align}
The full action is given by $S^\prime = S^\prime_1 + S^\prime_2 + S^\prime_\perp$

The new action is only bilinear in the Grassmannian fields, and hence, we can integrate them out from the partition function
\begin{align}
    \mathcal{Z} &= \int \mathcal{D}[c^*,c] e^{-S[c^*,c]} \nonumber \\
                &= \int \mathcal{D}[c^*,c]\mathcal{D}[\Psi^*,\Psi] e^{-S^\prime[c^*,c,\Psi^*,\Psi]} \nonumber \\
                & = \int \mathcal{D}[\Psi^*,\Psi]e^{-S_{\rm{eff}}[\Psi^*,\Psi]}.
\end{align}

In the last line, we have integrated out the Grassman fields and introduced a new quantity
\begin{align}
    S_{\rm{eff}}[\Psi^*,\Psi] := \beta_T \sum_l \frac{|\Psi_l|^2}{U} - \sum_{\vec{k},i\omega_n} \mathrm{Tr}\ln G^{-1}(\vec{k},i\omega_n,\Psi). 
    \label{effective}
\end{align}
In the above equation, $\vec{k} = \frac{\pi}{L}(n_x,n_y)$, where $n_x,n_y = -(L-1), \cdots, -1, 1, \cdots, (L-1)$, and $L$ is the size of the system in the x and y direction ($L$ must be even). The fermionic matsubara frequencies are given by: $\omega_n = \frac{(2m+1)\pi}{\beta}$, $m \in \mathbb{Z}$. The inverse Nambu Green's function can be written as, $G^{-1} = G_0^{-1} + (\Sigma(\Psi) + T)$, where,
{\small
\begin{equation}
    G_0^{-1} = \begin{pmatrix}
        -i\omega_n + \xi_{\vec{k}} & 0 & 0 & 0 \\
        0 & -i\omega_n - \xi_{\vec{k}} & 0 & 0 \\
        0 & 0 & -i\omega_n + \xi_{\vec{k}} & 0 \\
        0 & 0 & 0 & -i\omega_n - \xi_{\vec{k}}
    \end{pmatrix}
\end{equation}
}
\begin{equation}
    \Sigma(\Psi) = \begin{pmatrix}
        0 & \Psi_1 & 0 & 0 \\
        \Psi_1^* & 0 & 0 & 0 \\
        0 & 0 & 0 & \Psi_2 \\
        0 & 0 & \Psi_2^* & 0
    \end{pmatrix}
\end{equation}
\begin{equation}
    T = \begin{pmatrix}
        0 & 0 & -t_\perp & 0 \\
        0 & 0 & 0 & t_\perp \\
        -t_\perp & 0 & 0 & 0 \\
        0 & t_\perp & 0 & 0
    \end{pmatrix} \, \text{.}
\end{equation}
We treat $\Sigma$ and $T$ as perturbations, and expand: 
\begin{align}
    \ln G^{-1} &= \ln G_0^{-1} + \ln \big( 1+G_0(\Sigma + T) \big) \nonumber \\
               &= \ln G_0^{-1} + \sum_{m=1}^{\infty}(-1)^{m+1}\frac{\big(G_0(\Sigma + T)\big)^m}{m}
    \label{expansion}
\end{align}

The effective GL free energy $F[\Psi_1, \Psi_2] = \frac{S_{\rm{eff}}}{\beta_T}$:
\begin{align}
    F[\Psi_1, \Psi_2] = &\alpha\big(|\Psi_1|^2 + |\Psi_2|^2\big) + \beta \big(|\Psi_1|^4 + |\Psi_2|^4 \big) \nonumber \\ &+ J(\Psi_1 \Psi_2^* + \Psi_1^*\Psi_2) + \nonumber \\
    & K(|\nabla\Psi_1|^2 + |\nabla\Psi_1|^2)\cdots
\end{align}
The coefficient $\alpha$ comes from the second order term in Eq.~(\ref{expansion}), and $\beta$ and $J$ come from the fourth order term.
\begin{equation}
    \alpha = \frac{1}{U} - \sum_{\vec{k}}\frac{1}{\xi_{\vec{k}}} \tanh \Big( \frac{\beta_T \xi_{\vec{k}}}{2} \Big)
    \label{a}
\end{equation}

\begin{equation}
    \beta = \sum_{\vec{k}}\frac{1}{8\xi_{\vec{k}}^3}\Bigg[ \tanh \Big( \frac{\beta_T \xi_{\vec{k}}}{2}\Big) - \frac{1}{2}\beta_T\xi_{\vec{k}}\sech^2 \Big( \frac{\beta_T \xi_{\vec{k}}}{2} \Big) \Bigg]
    \label{b}
\end{equation}

\begin{equation}
    J = 2\beta t_\perp^2 \, \text{.}
    \label{J}
\end{equation}

To get the gradient term in the GL free energy we must allow $\Psi$ to slowly vary in space.

We can extract the temperature dependence of $\alpha$ from Eq.~(\ref{a}),
\begin{align}
    \sum_{\vec{k}}\frac{1}{\xi_{\vec{k}}} \tanh \Big( \frac{\beta_T \xi_{\vec{k}}}{2} \Big) &\approx N(0) \int_0^\Lambda d\xi \frac{1}{\xi_{\vec{k}}}\tanh \Big( \frac{\beta_T \xi_{\vec{k}}}{2} \Big) \nonumber \\
    & = N(0) \int_0^{\frac{\beta_T\Lambda}{2}} dx \frac{\tanh x}{x} \nonumber \\
    & = N(0) \ln(1.13\beta_T\Lambda) \, \text{.}
\end{align}
$\Lambda$ is just a cutoff and $N(0)$ is the density of states at the Fermi energy. We want $\alpha(T_c)=0$. Therefore, $\frac{1}{U} = N(0)\ln(1.13\beta_{T,c}\Lambda)$. Substituting, we get
\begin{align}
    \alpha = N(0)\ln\Big(\frac{T}{T_c}\Big)\approx \frac{N(0)}{T_c}(T-T_c)\, \text{.}
\end{align}

After allowing $\Psi$ to vary slowly and assuming that our dispersion relation is isotropic (which is true if we are close to the bottom of the band), we can also compute the coefficient $K$ as
\begin{align}
    K = 2&\sum_{\vec{k}}  \Big(v_{\vec{k}}^2 - \frac{1}{2}M_{\vec{k}}\xi_{\vec{k}}\Big) \nonumber \\
                        &\times \frac{1}{8\xi_{\vec{k}}^3}\Bigg[ \tanh \Big( \frac{\beta_T \xi_{\vec{k}}}{2}\Big) - \frac{1}{2}\beta_T\xi_{\vec{k}}\sech^2 \Big( \frac{\beta_T \xi_{\vec{k}}}{2} \Big) \Bigg],
\end{align}
where $v_{\vec{k}}:=\nabla_{\vec{k}}\xi_{\vec{k}}$, and $M_{\vec{k}}:=\nabla_{\vec{k}}^2\xi_{\vec{k}}$. We tabulate one set of lattice parameters and the corresponding GL coeffecients in Table~\ref{tab:GLcoeffs}.

\begin{table}[t]
\centering
\caption{Microscopic lattice parameters and corresponding Ginzburg--Landau (GL) coefficients.}
\label{tab:GLcoeffs}
\begin{tabular}{c c c c c | c c c c}
\hline\hline
\multicolumn{5}{c|}{Lattice parameters} & \multicolumn{4}{c}{GL coefficients} \\
\hline
$t$ & $\mu$ & $U$ & $t_\perp$ & $\delta T$ 
    & $a$ & $b$ & $K$ & $J$ \\
\hline
1.0 & $-0.02$ & 3.2 & 0.227 & 0.01 
    & $-0.938$ & 2.0306 & 0.09 & 0.209 \\
\hline\hline
\end{tabular}
\end{table}

\section{Sine Gordon equation for interlayer phase difference}

\label{app:SG}

The free energy in Eq.~(2) of the main text can be split into two parts:
\begin{equation}
    \Omega = \Omega_\psi + \Omega_\theta \, \text{,}
\end{equation}
where,
\begin{align}
    \label{omega_psi}
    \Omega_\psi(\psi_1,\psi_2) = \frac{1}{V_0}\int d^2\vec{r} \left[\sum_l \alpha_l\psi_l^2 + \frac{\beta}{2}\sum_l\psi_l^4 \right] \, \text{,}
\end{align}
\begin{align}
    \label{omega_theta}
    \Omega_\theta(&\psi_1, \psi_2, q,\phi) = \frac{1}{V_0}\int d^2\vec{r} \Bigg[ \frac{\psi_1^2}{2m}\left(q-q_{_{B}}+\frac{\nabla\phi}{2}\right)^2 \nonumber \\ &+ \frac{\psi_2^2}{2m}\left(q+q_{_{B}} - \frac{\nabla\phi}{2}\right)^2 -2J\psi_1\psi_2 \cos \phi\Bigg] \, \text{.}
\end{align}
$\Omega_\psi$ only depends on amplitudes $\psi_1$ and $\psi_2$, while $\Omega_\theta$ controls the phase fluctuations. We can separate the action $\Omega_\theta$ into 
\begin{eqnarray}
    \Omega_\theta = \Omega_1 + \Omega_2 + \Omega_3,
\end{eqnarray}
where
\begin{align}
    \label{omega1}
    \Omega_1 = \frac{1}{V_0}\int d^2\vec{r}\frac{1}{2m}\left(\psi_1^2 + \psi_2^2\right)q^2 \, \text{,}
\end{align}
\begin{align}
    \label{omega2}
    \Omega_2 = \frac{1}{V_0}\int d^2\vec{r}\Bigg[ &\frac{1}{2m}\left(\psi_1^2 + \psi_2^2\right)\left(\frac{\vec{\nabla}\phi}{2}-q_{_{B}} \right)^2 \nonumber \\ &- 2J\psi_1\psi_2 \cos\phi \Bigg] \, \text{,}
\end{align}
\begin{align}
    \label{omega3}
    \Omega_3 = \frac{1}{V_0}\int d^2\vec{r}\frac{q}{m}\left(\psi_1^2 -\psi_2^2\right)\left(\frac{\vec{\nabla}\phi}{2}-q_{_{B}} \right) \, \text{.}
\end{align}
Here $\Omega_1$ only depends on $q$ and is relevant to the calculation of supercurrent, $\Omega_2$ describes the vortex dynamics of relative phase $\phi$, and $\Omega_3$ describes the coupling between the momentum $q$ and the relative phase $\phi$ and can also contributes to supercurrent. $\Omega_3$ is crucial for the coupling between the supercurrent and the interlayer vortices, which requires to break inversion symmetry between two layers since it includes the difference $\psi^2_1-\psi^2_2$ between the SC order parameters in two layers.

We first consider Eq.~(\ref{omega2}) at fixed $\psi_1$ and $\psi_2$. After redefining $m^\prime =2m\psi_0^2/(\psi_1^2+\psi_2^2)$ and $J^\prime =\psi_1\psi_2J/\psi_0^2$, with $\psi_0 = (\psi_1+\psi_2)/2$, we get the sine-Gordon type of action for $\Omega_2$,
\begin{align}
    \Omega_2 =\frac{\psi_0^2}{L_x}\int dx\Big[\frac{1}{m^\prime}\big(\tfrac{1}{2}\partial_x\phi-q_{_{B}}\big)^2 - 2J^\prime\cos\phi\Big]
\end{align}

Minimizing $\Omega_2$ leads to,
\begin{align}
    \delta\Omega_2 &= \frac{1}{L_x}\int dx\Big[\frac{1}{m^\prime}\big(\tfrac{1}{2}\partial_x\phi-q_{_{B}}\big)\partial_x(\delta\phi) +2J^\prime\sin\phi \delta\phi \Big] \nonumber\\
                      &=\frac{1}{L_x}\int dx\Big[\frac{1}{2m^\prime}(-\partial^2_x\phi) + 2J^\prime\sin\phi \Big]\delta\phi   \, \text{.}
\end{align}
In the above steps, we used integral by parts, and dropped the boundary term. With $\delta\Omega_2 = 0$, we obtain the sine-Gordon equation:
\begin{equation}
    \label{SG1}
    -\frac{1}{2m^\prime}\partial_x^2\phi + 2J^\prime\sin\phi = 0 \, \text{.}
\end{equation}
With the rescaling, $w=(4J^\prime m^\prime)^{1/2}x$, Eq.~(\ref{SG1}) becomes,
\begin{equation}
    \label{SG2}
    \partial_w^2\phi = \sin\phi \, \text{.}
\end{equation}

Equation (\ref{SG2}) can be rewritten as $$\partial_w\big[\tfrac{1}{2}(\partial_w\phi)^2+\cos\phi\big] = 0 \, \text{,}$$ which gives,
\begin{align}
            &\partial_w\phi(w) = \pm\sqrt{C_0 - 2\cos\phi}\nonumber \\
    \implies&dw = \pm\frac{d\phi}{\sqrt{C_0 - 2\cos\phi}} \nonumber \\
    \implies&\int_{w_0}^{w}dw = \pm\int_{\phi(w_0)}^{\phi(w)}\frac{d\phi^\prime}{\sqrt{C_0 - 2\cos\phi^\prime}} \, \text{,} \nonumber
\end{align}
where $C_0$ is a constant. By defining $\tilde{\phi} = \phi+\pi$, we have
\begin{align}
    w-w_0 &= \pm\int_{\tilde{\phi}(w_0)}^{\tilde{\phi}(w)}\frac{d\tilde{\phi}^\prime}{\sqrt{C_0 + 2\cos\tilde{\phi}^\prime}} \nonumber\\
          &= \pm \frac{1}{\sqrt{C_0+2}} \int_{\tilde{\phi}(w_0)}^{\tilde{\phi}(w)} \frac{d\tilde{\phi}^\prime}{\sqrt{1-\frac{4}{C_0+2}\sin^2\Big(\frac{\tilde{\phi}^\prime}{2}\Big)}} \nonumber
\end{align}
We choose the initial condition $\tilde{\phi}(w_0) = 0 \implies \phi(w_0) = -\pi$, and change integration variables to $\theta = \tilde{\phi}/2$, which gives
\begin{equation}
    \label{Ei1}
    w-w_0 = \pm k\int_0^\theta\frac{d\theta^\prime}{\sqrt{1-k^2\sin^2\theta^\prime}} \, \text{,}
\end{equation}
where, $k=\frac{2}{\sqrt{C_0+2}}$. The right hand side of the above equation is the incomplete elliptic integral of the first kind, provided $k\in[0,1]$, or $C_0\geq2$. The limiting cases of $k=1$ or $C_0=2$ correspond zero interlayer vortices. As $k$ decreases from $1$ or $C_0$ increases from $2$, the number of vortices in the system increases. The ground state is obtained by minimizing the total free energy (with all terms) with respect to $k$, $\psi_1$, and $\psi_2$. 

The solution to Eq.~(\ref{Ei1}) is given by
\begin{align}
    \label{LM}
    \theta &= \am\Big(\pm\frac{1}{k}(w-w_0),k^2 \Big) \nonumber \\
    \phi(w) &= 2\am\Big(\pm\frac{1}{k}(w-w_0),k^2 \Big) - \pi \, \text{,}
\end{align}
where, $\am(u,m)$ is the Jacobi-amplitude function~\cite{abramowitz1965handbook, Scott1973}. This gives the local minima of $\Omega_2$ for a fixed $\psi_1$ and $\psi_2$,
\begin{align}
    \Omega^\prime_2 &= \frac{\psi_0^2}{L_x}\int dx\Big[\frac{1}{m^\prime}\big(\tfrac{1}{2}\partial_x\phi-q_{_{B}}\big)^2 - 2J^\prime\cos\phi\Big] \nonumber \\
                       &= \frac{\psi_0^2}{m^\prime L_x}\int dx\Big[\big(\tfrac{1}{2}\partial_x\tilde{\phi}-q_{_{B}}\big)^2 + 2J^\prime m^\prime\cos\tilde{\phi}\Big] \nonumber
\end{align}
We can substitute Eq.~(\ref{LM}) in the expression for $\Omega_2$. Choosing $w_0=0$, and after some algebra, we get,
\begin{align}
    \Omega^\prime_2 = \frac{\psi_0^2}{m^\prime}\Bigg[& q_{_{B}}^2 -C_0J^\prime m^\prime + \nonumber \\ &\frac{2}{L_x}\sqrt{J^\prime m^\prime(C_0+2)}\varepsilon(\sqrt{J^\prime m^\prime(C_0+2)}L_x,k^2) \nonumber \\ & \mp \frac{2q_{_{B}}}{L_x}\am(\sqrt{J^\prime m^\prime(C_0+2)}L_x,k^2) \Bigg] \, \text{,}
\end{align}
where, $\varepsilon(u,m)$ is the Jacobi epsilon function~\cite{abramowitz1965handbook, Scott1973}. Choosing the negative sign in Eq.~(\ref{LM}) gives us the positive sign in the above equation and vice versa. For $q_{_{B}}>0$, choosing the positive sign in Eq.~(\ref{LM}) always gives us the lower energy solution and vice versa. For $q_{_{B}}=0$, it may seem there are two degenerate states corresponding to the two distinct solutions in Eq.~(\ref{LM}), but then the minimum occurs at $k=1$, and the amplitude function is zero, giving us only one state.
We have $\vec{\nabla}\xi = 0$ in the FFLO state (in the gauge we have chosen in the main text). The free energy contribution from the overall phase is
\begin{equation}
    \Omega_1 = \frac{1}{2m}\big(\psi_1^2+\psi_2^2 \big)q^2 \, \text{.}
\end{equation}
The coupling term $\Omega_3$ between $\xi$ and $\phi$ is given by,
\begin{align}
    \Omega_3 &= \frac{\psi_1^2-\psi_2^2}{m}\frac{q}{L_x}\int dx\big(\tfrac{\partial_x\phi}{2} - q_{_{B}}\big) \nonumber \\
                     &= \frac{\psi_1^2-\psi_2^2}{m}q\Big(\frac{1}{L_x}\am(\sqrt{J^\prime m^\prime(C_0+2)}L_x,k^2) - q_{_{B}} \Big) \, \text{.}
\end{align}
To consider the thermodynamic limit $L_x\rightarrow\infty$, we use the asymptotic behavior of $\varepsilon(u,m)$ and $\am(u,m)$,
$$\lim_{L_x \to \infty} \frac{1}{L_x}\varepsilon(aL_x,k^2) = a\frac{E(k^2)}{K(k^2)}$$ \, \text{.}
\begin{align}
    \lim_{L_x \to \infty} \frac{1}{L_x}\am(aL_x,k^2) = a\frac{\pi}{2K(k^2)} \nonumber \, \text{,}
\end{align}
where, $K(m)$ and $E(m)$ are the complete elliptic integrals of the first and second kind, respectively.

Using the above expressions and substituting the expressions for $J^\prime$, $m^\prime$, and $C_0$ we obtain,
\begin{align}
    \label{omega2_sg}
    \Omega_2 =& \frac{q_{_{B}}^2}{2m}\big( \psi_1^2 + \psi_2^2 \big) + A(k)\psi_1\psi_2 \nonumber \\  &- B(k) q_{_{B}}\left( \frac{\psi_1\psi_2}{2}(\psi_1^2+\psi_2^2) \right)^{1/2} \, \text{,}
\end{align}
\begin{equation}
    \label{omega3_sg}
    \Omega_3 =  \frac{\psi_1^2-\psi_2^2}{2} q \Bigg[B(k) \left(\frac{2\psi_1\psi_2}{\psi_1^2+\psi_2^2}\right)^{1/2} - \frac{2q_{_{B}}}{m}\Bigg] \, \text{,}
\end{equation}
where, 
\begin{equation}
    \label{Ak}
    A(k) = 2J(1-2/k^2 + (4/k^2)(E(k^2)/K(k^2))) \, \text{,}
\end{equation}
\begin{equation}
    \label{Bk}
    B(k) = (2/k)(J/m)^{1/2}(\pi/K(k^2)) \, \text{.}
\end{equation}
The other terms remain unchanged.

For a fixed $q$ we minimize the free energy with respect to $k$, $\psi_1$, and $\psi_2$, and obtain the ground state. When the minimum occurs at $k=1$, there are no vortices in the system and we call this the BCS state. When the minimum occurs at $k<1$, there are interlayer vortices, and the system is in the FFLO state. The current through the system is then determined from Eq.~(5) in the main text.

\section{The low magnetic field limit}

\label{app:lowmag}

In this section we analyze the behaviour of $\eta$ in the low $q_{_{B}}$ regime. We know from numerical calculations that in this regime the system prefers to be in the BCS state ($\nabla\phi = 0$). So in this section, we may restrict ourselves to the BCS saddle and rewrite the free energy as
\begin{eqnarray}
    \label{omegabcs}
    \Omega_{\rm{BCS}} = &\sum_{l=1.2} \left[ \alpha_l\psi_1^2 +\frac{\beta}{2}\psi_l^4 + \frac{\psi_l^2}{2m} (q+(-1)^l q_{_{B}})^2 \right] \nonumber \\ &-2J\psi_1\psi_2 \, \text{.}
\end{eqnarray}
The current is given by,
\begin{eqnarray}
    I_x = \frac{2e}{m} \left[ (\psi_1^2 + \psi_2^2)q - (\psi_1^2 - \psi_2^2)q_{_{B}} \right] \text{.}
\end{eqnarray}
We define the following quantities,
\begin{eqnarray}
    a_1 (q,q_{_{B}}) = \alpha_1 + \frac{(q-q_{_{B}})^2}{2m} \text{,} \\
    a_2(q,q_{_{B}}) = \alpha_2 + \frac{(q+q_{_{B}})^2}{2m} \text{.}
\end{eqnarray}
Using these newly defined quantities, Eq.~(\ref{omegabcs}) can be rewritten as
\begin{equation}
    \label{omegabcs2}
    \Omega_{\rm{BCS}} = a_1\psi_1^2 + \frac{\beta}{2}\psi_1^2 +a_2\psi_2^2 + \frac{\beta}{2}\psi_2^2 -2J\psi_1\psi_2
\end{equation}
The BCS saddle is determined by minimizing the expression in Eq.~(\ref{omegabcs2}) with respect to $\psi_1$ and $\psi_2$, which yields the following coupled equations,
\begin{eqnarray}
    a_1 (q,q_{_{B}}) \psi_1 + \beta \psi_1^3 - J\psi_2 = 0 , \label{scbcs1}\\
    a_2 (q,q_{_{B}}) \psi_2 + \beta \psi_2^3 - J\psi_1 = 0 . \label{scbcs2}
\end{eqnarray}
In the main text, we defined the critical currents $I_c^\pm$ at a fixed $q_{_{B}}$ as the magnitude of the maximum and minimum value of $I_x$ as a function of $q$ respectively. Let $q_\pm(q_{_{B}})$ be the values of $q$ for which these extremum values are achieved. That is,
\begin{eqnarray}
    I_c^+(q_{_{B}}) = I_x(q_+(q_{_{B}}),q_{_{B}}), \\
    I_c^-(q_{_{B}}) = I_x(q_-(q_{_{B}}),q_{_{B}}).
\end{eqnarray}
We wish to obtain a small $q_{_{B}}$ expansion for the critical currents. Let $q_+(0) = q_0$, and from symmetry of $\Omega_{\rm{BCS}}$, we know that $q_-(0) = -q_0$. Let $I_c^+(0) = I_0 = -I_c^-(0)$. Then, for small $q_{_{B}}$,
\begin{equation}
    I_c^+(q_{_{B}}) = I_0 + \partial_q I_x|_{(q_0,0)} \delta q_0 + \partial_{q_{_{B}}}I_x|_{(q_0,0)}q_{_{B}} + O(q_{_{B}}^2).
\end{equation}
But, by definition, $\partial_q I_x|_{(q_0,0)}=0$. Therefore,
\begin{equation}
    I_c^+(q_{_{B}}) = I_0 +  \partial_{q_{_{B}}}I_x|_{(q_0,0)}q_{_{B}} + O(q_{_{B}}^2).
\end{equation}
Similarly,
\begin{equation}
    I_c^-(q_{_{B}}) = I_0 -  \partial_{q_{_{B}}}I_x|_{(-q_0,0)}q_{_{B}} + O(q_{_{B}}^2).
\end{equation}
Now, $\Omega_{\rm{BCS}}(q,q_{_{B}}) = \Omega_{\rm{BCS}}(-q,-q_{_{B}})$. Since $I_x = 2e\partial_q\Omega_{\rm{BCS}}$, $\partial_{q_{_{B}}}I_x(q_0,0) = \partial_{q_{_{B}}}I_x(-q_0,0)$. Let us call this quantity $I_0^\prime$. Using the formula for the diode efficiency from Eq.~(1) in the main text, we get,
\begin{equation}
    \eta(q_{_{B}}) \approx q_{_{B}}\frac{I_0^\prime}{I_0}\propto q_{_{B}}
\end{equation}
To determine the exact slope, one needs to solve the system of Eqs.~(\ref{scbcs1}) and (\ref{scbcs2}) for $q_{_{B}}=0$ and arbitrary values of $q$, and obtain the value of the currents. Since it is a coupled system of non linear equations, this is best done numerically. But we have shown that close to zero magnetic field, the leading contribution to $\eta$ in linear in $q_{_{B}}$.

\section{The high magnetic field limit}

\label{app:highmag1}

In this section, we carry out the minimization with respect to $k$ in the high magnetic field limit, and see what happens to the diode efficiency, $\eta$. We already derived the full free energy in the sine-Gordon saddle in the previous section, given by Eqns.~(\ref{omega_psi}),(\ref{omega1}),(\ref{omega2_sg}),(\ref{omega3_sg}). Below, we define
\begin{align}
    S = \psi_1^2+\psi_2^2 \, \text{,}
    \label{S}
\end{align}
\begin{align}
    D = \psi_1^2-\psi_2^2 \, \text{,}
    \label{D}
\end{align}
\begin{align}
    P=\psi_1\psi_2 \, \text{.}
\end{align}
Using these quantities, we rewrite the action as
\begin{align}
    \Omega_1 = \frac{q^2}{2m}S \, \text{,}
\end{align}
\begin{align}
    \Omega_2 = \frac{q_{_{B}}^2}{2m}S + A(k)P - B(k)q_{_{B}}\Big(\frac{PS}{2} \Big)^{1/2} \, \text{,}
\end{align}
\begin{align}
    \Omega_3 = \frac{D}{2}q\Bigg(B(k)\Big(\frac{2P}{S} \Big)^{1/2} -\frac{2q_{_{B}}}{m} \Bigg) \, \text{,}
\end{align}
where $A(k)$ and $B(k)$ has been defined in the previous section. We minimize the action with respect to $k$ with fixed $\psi_1$ and $\psi_2$. By defining quantities $u = (2P/S)^{1/2}$ and $\epsilon = (qD)/S$, $\partial\Omega/\partial k = 0$ gives
\begin{align}
    uA'(k) = (q_{_{B}} - \epsilon)B'(k) \, \text{.}
    \label{kmin}
\end{align}
At high $q_{_{B}}$, $k$ should be small. So we use the small $k$ asymptotic behavior of the elliptic integrals,
\begin{align}
    A(k) \approx 2J \Big(\frac{2}{k^2} - 1 -\frac{k^2}{4} + O(k^4)\Big)
\end{align}
\begin{align}
    B(k) = \bigg(\frac{J}{m}\bigg)^{1/2} \bigg(\frac{4}{k} - k - \frac{5k^3}{16} + O(k^5)\bigg)
\end{align}
Substituting these expressions in Eq.(~\ref{kmin}), and keeping only leading powers of $k$, we find that the free energy $\Omega$ is stationary with respect to $k$ when:
\begin{eqnarray}
    k^* = \frac{2\sqrt{Jm}}{q_{_{B}} - \epsilon}u
    \label{kstar}
\end{eqnarray}
Before we substitute this in $\Omega$ to get the effective free energy, let us first see how $\phi(x)$ looks like in this limit. We have have obtained the expression for $\phi(x)$ in Eq.~\ref{LM}. At large $q_{_{B}}$, $k^2 \approx 0$. So,
\begin{align}
    \phi^*(x) &\approx 2\am\bigg(\frac{\sqrt{4J^\prime m^\prime}}{k^*}x,0 \bigg) - \pi \nonumber \\
              &=2(q_{_{B}}-\epsilon)\bigg(\frac{J^\prime m^\prime}{Jm}\bigg)^{1/2} \frac{x}{u} - \pi \nonumber \\
              &=2(q_{_{B}}-\epsilon)x - \pi \, \text{,} \label{phistar}
\end{align}
where, we have defined $\phi(x)|_{k=k^*} = \phi^*(x)$. So, when $q_{_{B}} \gg q$, $\phi^*(x) \approx2q_{_{B}}x - \pi$. 

To understand how the diode efficiency ($\eta$) behaves in this limit, we substitute $k=k^*$ given by Eq.~(\ref{kstar}) and expand in orders of $(1/q_{_{B}})$. After some algebra, the free energy is given by
\begin{widetext}   
\begin{equation} 
    \Omega_{\rm{eff}}(\psi_1,\psi_2,q) = \Omega_\psi - \frac{J^2m}{q_{_{B}}^2}\frac{4\psi_1^2\psi_2^2}{\psi_1^2 + \psi_2^2} + \frac{q^2}{2m}\left( \psi_1^2 +\psi_2^2 - \frac{(\psi_1^2 - \psi_2^2)^2}{\psi_1^2+\psi_2^2} - \frac{4Jm}{q_{_{B}}^2} \left( \frac{\psi_1^2 - \psi_2^2}{\psi_1^2 + \psi_2^2} \right)^2\psi_1\psi_2 \right)  + O\left(\frac{1}{q_{_{B}}^3}\right) \, \text{,}
    \label{om_eff}
\end{equation}
\end{widetext}
where, $\Omega_\psi$ is given by Eq.~(\ref{omega_psi}). From Eq.~(\ref{om_eff}), we note that up to $O(1/q_{_{B}}^2)$, no odd powers of $q$ survive in $\Omega_{\rm{eff}}$. This means that at high $q_{_{B}}$, $\Omega_{\rm{eff}}(q) \approx\Omega_{\rm{eff}}(-q)$. Since the current through the superconductor in given by $I = 2e\partial\Omega_{\rm{eff}}/\partial q$, this implies that the currents are reciprocal in this limit, and $\eta\rightarrow0$ as $q_{_{B}}\rightarrow\infty$. 

\section{Effective decoupling of layers in the high $B$ limit}

\label{app:highmag}
In this section, we wish to recast the free energy in terms of dimensionless quantities to show that the $q_{_{B}}\rightarrow\infty$ limit is equivalent to the $J\rightarrow 0$ limit, where the two layers are decoupled. We consider the $q=0$ case first. The angular part of free energy, $\Omega_\theta$ is
\begin{align}
    \label{omega_theta_q0}
    \Omega_\theta(\psi_1, \psi_2, q,\phi) = & \frac{1}{L_x}\int dx \Bigg[ \frac{\psi_1^2+\psi_2^2}{2m}\left(\frac{\partial_x\phi}{2}-q_{_{B}}\right)^2\nonumber \\ &-2J\psi_1\psi_2 \cos \phi\Bigg] \, \text{.}
\end{align}
With $w = (4J^\prime m^\prime)^{1/2}x$, $m^\prime =2m\psi_0^2/(\psi_1^2+\psi_2^2)$, $J^\prime =\psi_1\psi_2J/\psi_0^2$, and $\psi_0 = (\psi_1+\psi_2)/2$, we have shown earlier that this reduces the Euler-Lagrange equation to the standard sine-Gordon form with no dimension-ful quantities, given by Eq.~(\ref{SG2}). $\Omega_\theta$ becomes
\begin{align}
    \Omega_\theta &= \frac{\psi_0^2}{L_x}\int dx \Bigg[ \frac{1}{m^\prime} \left(\frac{\partial_x\phi}{2}-q_{_{B}}\right)^2 - 2J^\prime \cos \phi\Bigg] \nonumber \\
                  &=\Gamma\int dw \big[ (\partial_w\phi - \delta)^2 - 2\cos\phi\big] \, \text{,}
\end{align}
where
\begin{align}
    \Gamma = \frac{1}{2\sqrt{2}L_x}\left( \frac{J}{m}\psi_1\psi_2(\psi_1^2 + \psi_2^2) \right)^{1/2} \, \text{.}
\end{align}
The action in the integral is only controlled by the dimensionless quantity
\begin{align}
    \delta = \frac{q_{_{B}}}{\sqrt{Jm}}\left( \frac{\psi_1^2 + \psi_2^2}{2\psi_1\psi_2} \right)^{1/2} \, \text{.}
\end{align}
For high values of $\delta$, the system would tend to prefer a non zero $\partial_w\phi$ to reduce the magnitude of the first term in the integral, and for small values of $\delta$, the system would prefer $\phi=0$ to minimize the second term. So the competition between the BCS and the FFLO states in controlled entirely by the parameter $\delta \propto q_{_{B}}/\sqrt{J}$, Therefore, $q_{_{B}} \rightarrow \infty$ limit is equivalent to the $J\rightarrow 0$ limit, where the two layers are effectively decoupled. In this limit, since the two layers behave independently, the diode efficiency, $\eta=0$.

If we bring back the terms containing $q$, 
\begin{align}
    \Omega_\theta = \frac{\psi_0^2}{L_x}\int dx \Bigg[ \frac{1}{m^\prime} \left(\frac{\partial_x\phi}{2}-q_{_{B}}\right)^2 - 2J^\prime \cos \phi + \frac{q^2}{m^\prime} \nonumber \\ \frac{2q}{m^\prime}\frac{\psi_1^2-\psi_2^2}{\psi_1^2+\psi_2^2}\left(\frac{\partial_x\phi}{2}-q_{_{B}}\right)  \Bigg]
\end{align}
Note that the third term is constant in $\phi$ and the fourth term is a total derivative, so they do not affect the Sine-Gordon action. After doing a similar rescaling, and defining a dimensionless quantity
\begin{align}
    \delta_q = \frac{q}{\sqrt{Jm}}\frac{\psi_1^2-\psi_2^2}{(2\psi_1\psi_2(\psi_1^2+\psi_2^2))^{1/2}} \, \text{,}
\end{align}
we get the free energy
\begin{align}
    \Omega_\theta = \Gamma \int dw \Bigg[ (\partial_w\phi -\delta)^2 - 2\cos\phi +\delta_q^2 \left( \frac{\psi_1^2+\psi_2^2}{\psi_1^2-\psi_2^2} \right)^2 \nonumber \\ +2\delta_q(\partial_w\phi -\delta) \Bigg] \, \text{.}
\end{align}
The final two terms do not contribute to the Sine-Gordon equation, as the third term is a constant in $\phi$ and the final terms is total derivative in $\phi$. 

\section{The vortex density and its asymptotic limit}

\label{app:asymptotic}
In this section, we derive the vortex density of the bilayer system in two limiting regimes. We consider first the high magnetic-field limit, and then the regime close to the critical magnetic field at which vortices enter the system. A vortex in this system is defined as whenever the relative phase between the two layers changes by $2\pi$. From the thermodynamic limit of the Jacobi amplitude function, the relative phase from Eq.~\ref{LM} is given by:
\begin{equation}
    \label{relp}
    \phi(x) = \frac{2\pi\sqrt{J^\prime m^\prime}}{kK(k^2)}x
\end{equation}
Hence, from this, the linear density of vortices in the system is given by
\begin{equation}
    \label{rho_def}
    \rho = \frac{\sqrt{J^\prime m^\prime}}{kK(k^2)} \, \text{.}
\end{equation}

We have already shown, that at high magnetic fields, $\phi(x) = 2q_{_{B}}x-\pi$ in Eq.~(\ref{phistar}), which gives the asymptotic value of $\rho \rightarrow q_{_{B}}/\pi$. We wish to determine how $\rho$ approches this limit. We have shown that to leading order in $q_{_{B}}$, the value of $k$ that minimizes the free energy is given by Eq.~\ref{kstar}. To determine higher order corrections to $k^\star$, we define:
\begin{equation}
    F(k) = k^3 \big( uA^\prime (k) - (q_{_{B}}-\epsilon)B^\prime (k) \big)
    \label{F}
\end{equation}
From Eq.~(\ref{kmin}), we see that the minimization condition $\partial\Omega/\partial k = 0$ is equivalent to $F(k) = 0$. Let us substitute $k = k^\star + \delta k$ in this equation. Expanding around $k=k^\star$ yields $F(k^*) + F^\prime(k^*)\delta k = 0$, so $\delta k = -(F(k^*)/F^\prime(k^*))$. Keeping only the lowest order correction, we find $\delta k = -{k^\star}^3/4$. Substituting $\delta k$ in Eq.~(\ref{rho_def}), we find the asymptotic behaviour of the density of vortices at high magnetic fields,
\begin{equation}
    \label{rho_high}
    \rho_{\rm{high}} = \frac{q_{_{B}}}{\pi} - \frac{u^4}{4\pi}\frac{(Jm)^2}{q_{_{B}}^3} \, \text{.}
\end{equation}

We next investigate the other limit when the magnetic field is only slightly higher that the critical field at which vortices start entering. Before we do that, we can compute $A^\prime(k)$ and $B^\prime (k)$ as follows from Eqns.~(\ref{Ak}) and (\ref{Bk}),
\begin{eqnarray}
    A^\prime(k) =& -\frac{8J}{k^3(1-k^2)}\frac{E^2(k^2)}{K^2(k^2)} \\
    B^\prime(k) =& -2\pi \sqrt{\frac{J}{m}} \frac{E(k^2)}{k^2(1-k^2)K^2(k^2)} 
\end{eqnarray}
This simplifies the minimization condition in Eq.~(\ref{kmin}) as
\begin{equation}
    q_{_{B}}-\epsilon = \frac{4u}{\pi}\sqrt{Jm}\frac{E(k^2)}{k}
    \label{kmin_new}
\end{equation}
To compute the asymptotic behavior in the sparse vortex limit, we choose $k = 1-\delta$, with $\delta \ll 1$. It is useful to define the quantity $k^\prime = \sqrt{1-k^2} \approx \sqrt{2\delta}$. For simplicity, we can also drop the $\epsilon$. The asymptotic expansion for $E(k^2)$ in the $k \rightarrow 1^-$ limit is
\begin{equation}
    \label{Ekasymp}
    E(k^2) = 1 + \frac{{k^\prime}^2}{2} \Bigg( \ln \Big(\frac{4}{k^\prime}\Big) - \frac{1}{2}  \Bigg) + \cdots \,\text{.}
\end{equation}
Substituting Eq.~(\ref{Ekasymp}) in Eq.~(\ref{kmin_new}), and keeping only leading contributions, we get
\begin{eqnarray}
    q_{_{B}} = \frac{4u}{\pi}\sqrt{Jm} \left( 1 + \delta \ln \left( \frac{4}{\sqrt{2\delta}} \right) \right) \,\text{.}
    \label{lowB}
\end{eqnarray}

In the limit $\delta \rightarrow 0^+$, $q_{_{B}} \rightarrow \frac{4u}{\pi}\sqrt{Jm}$, which is, therefore, the critical magnetic field $q_{_{B,c}}$. Solving Eq.~(\ref{lowB}) for $\delta$, and using the fact $\delta$ is small, we get
\begin{eqnarray}
    \delta = \frac{r}{\ln \left( \frac{4}{r} \right)}\,\text{,}
\end{eqnarray}
where, we have used $r = (q_{_{B}}/q_{_{B,c}})-1$. Substituting $k = 1-\delta$ in Eq.~(\ref{rho_def}), we get the low vortex density behavior
\begin{eqnarray}
    \rho_{\rm{low}} = \frac{2u\sqrt{Jm}}{\ln\left( \frac{8}{\left( \frac{q_{_{B}}}{q_{_{B,c}}} -1 \right)} \right)} \,\text{.}
    \label{rho_low}
\end{eqnarray}
From Eq.(\ref{rho_low}), one can see the vortex density $\rho$ approaches zero as $q_{_{B}} \rightarrow q_{_{B,c}}^+$, so we would expect the BCS-FFLO transition is continuous in our system.

\bibliographystyle{unsrtnat}
\bibliography{Schirmer.bib}